\documentclass[]{scrartcl}
\usepackage{graphics}
\usepackage[english]{babel}
\usepackage{amsmath} 
\usepackage{amsfonts}
\usepackage{latexsym}
\usepackage{cuted}
\usepackage{graphicx}
\usepackage{graphics}
\usepackage{float}
\usepackage{breqn}
\usepackage{amssymb}
\usepackage{authblk}
\usepackage{multirow}
\usepackage{makeidx}
\usepackage{dblfloatfix}
\usepackage[section]{placeins}

\PassOptionsToPackage{hyphens}{url}
\usepackage{hyperref}
\usepackage{url}
\usepackage{graphics}
\usepackage[english]{babel}
\usepackage{amsmath} 
\usepackage{amsfonts}
\usepackage{latexsym}
\usepackage{graphicx}
\usepackage{graphics}
\usepackage{float}
\usepackage{hyperref}
\usepackage{amssymb}
\usepackage{authblk}
\usepackage{multirow}
\usepackage{makeidx}
\usepackage{dblfloatfix}
\usepackage[section]{placeins}
\usepackage{url}
\usepackage{flushend}
\begin{document}
\title{A new laser-ranged satellite
for General Relativity and Space Geodesy
}
\subtitle{I. An introduction to the LARES2 space experiment.
}

\author[1,2]{Ignazio Ciufolini\thanks{ignazio.ciufolini@unisalento.it}}
\author[3]{Antonio Paolozzi}
\author[4]{Erricos C. Pavlis}
\author[3]{Giampiero Sindoni}
\author[5]{Rolf Koenig}
\author[6]{John Ries}
\author[7]{Richard Matzner}
\author[8]{Vahe Gurzadyan}
\author[9]{ Roger Penrose}
\author[10]{ David Rubincam}
\author[2,3]{Claudio Paris}

\affil[1]{\footnotesize  Dip. Ingegneria dell'Innovazione, Universit\`a del Salento, Lecce, Italy}
\affil[2]{Museo della fisica e Centro studi e ricerche Enrico Fermi, Rome, Italy}
\affil[3]{Scuola di Ingegneria Aerospaziale, Sapienza Universit\`a di Roma, Italy}
\affil[4]{Joint Center for Earth Systems Technology (JCET), University of Maryland, Baltimore County, USA}
\affil[5]{Helmholtz Centre Potsdam, GFZ German Research Centre for Geosciences, Potsdam, Germany}
\affil[6]{Center for Space Research, University of Texas at Austin, Austin, USA}
\affil[7]{Theory Center, University of Texas at Austin, Austin, USA}

\affil[8]{Center for Cosmology and Astrophysics, Alikhanian National Laboratory and Yerevan State University, Yerevan, Armenia}

\affil[9]{Mathematical Institute, University of Oxford, Oxford, UK}
\affil[10]{NASA Goddard Space Flight Center, Greenbelt, Maryland, USA}

\renewcommand\Authands{ and }

\date{}
\maketitle
\abstract{
 We introduce the LARES 2 space experiment recently approved by the Italian Space Agency (ASI). The LARES 2 satellite is planned for a launch in 2019 with the new VEGA C launch vehicle of ASI, ESA and ELV. The orbital analysis of LARES 2 experiment will be carried out by our international science team of experts in General Relativity, theoretical physics, space geodesy and aerospace engineering. The main objectives of the LARES 2 experiment are gravitational and fundamental physics, including accurate measurements of General Relativity, in particular a test of frame-dragging aimed at achieving an accuracy of a few parts in a thousand, i.e., aimed at improving by about an order of magnitude the present state-of-the-art and forthcoming tests of this general relativistic phenomenon. LARES 2 will also achieve  determinations in space geodesy. LARES 2 is an improved version of the LAGEOS 3 experiment, proposed in 1984 to measure frame-dragging and analyzed in 1989 by a joint ASI and NASA study.

} 
\section{Introduction: LARES 2 and frame-dragging}
\label{intro}
The main objective of the LARES 2 space experiment is a measurement of General relativitic frame-dragging \cite{bib1,bib2,bib3,bib4,bib5,bib5bis,bib6} with an accuracy of approximately 0.2\%, approaching parts in a thousand as justified both in terms of the error budget calculated below and on the basis of Monte Carlo simulations and covariance analyses presented in the next paper (Paper II: \cite{bib7}). Thus LARES 2 will basically improve by about an order of magnitude the best measurements of frame-dragging that can be achieved by the LARES space mission only (without LARES 2) \cite{bib8,bib9,bib10,bib11,bib12,bib13,bib14,bib15,bib16} In addition LARES 2 will provide important contributions in Space Geodesy and Geodynamics, as described below.

Frame-dragging is the change of orientation of the axes of local inertial frames where the equivalence principle, at the basis of General Relativity, holds \cite{bib6}. The equivalence principle states that in local freely-falling frames all the laws of physics are the laws of Special Relativity \cite{bib3,bib5,bib19}. In other words, in local freely-falling frames it is possible to eliminate the effects of the gravitational field in the sense of making them arbitrarily small in a sufficiently small spacetime neighborhood of a freely-falling frame. The axes of these local inertial frames are determined in General Relativity by gyroscopes. However, the gyroscopes are not fixed with respect to the ``distant stars'' (as in classical mechanics) but they are dragged by the mass-energy currents, such as the rotation of a nearby body. This effect -- also called the Lense-Thirring effect [1,2,6] -- is formally similar (in the weak field and slow motion approximation) to the change of orientation of a magnetic dipole due to the magnetic field generated by an electric current. Frame-dragging is very tiny (but observable, as we discuss below) around the rotating Earth but has huge effects around rotating black holes. Thus frame-dragging plays a key role in the dynamics of a body and of the accretion disk around rotating black holes and in the dynamics of jets in active galactic nuclei \cite{bib4}. The first direct observation in September 2016 of gravitational waves with the two LIGO laser interferometers opened the era of gravitational wave astronomy \cite{bib22}. The observation captured the spectacular collision of two black holes forming, as predicted by General Relativity, a single rotating black hole with an enormous release of energy in the form of gravitational waves. There is no evidence that the initial black holes in the 2016 event had significant spin, but mutual frame-dragging of colliding initially rotating blacks holes will be an important factor in the analysis of the signal of the gravitational waves emitted by such systems, when they are detected. The accurate measurement of frame-dragging will thus be an important ingredient in the study of the coalescence of rotating black holes.

With LARES 2 we could further accurately test these fundamental physics theories. Frame-dragging of a rotating
planet produces on its satellites a precession of the orbital
angular momentum of a satellite due to the angular momentum of the central body, usually described as the change
of orientation of its nodal line (the intersection of its orbital plane with the equatorial plane of the planet). The
Lense-Thirring drag of a satellite orbital plane and node by the angular momentum of the central body has a sense of
rotation that is the same of the rotation of the central body (see Figure 1).
A number of gravitational theories alternative to General Relativity (the so-called f(R) theories) have been proposed
in the attempt to elucidate the profound mystery of dark energy and quintessence and so to explain the accelerated
expansion of the universe. Some of these theories (such as Chern-Simons gravity and string theories equivalent to
it) predict an outcome for frame-dragging different from General Relativity, even around the rotating Earth [19].

\begin{figure}
\centering
 \includegraphics[width=0.640\textwidth]{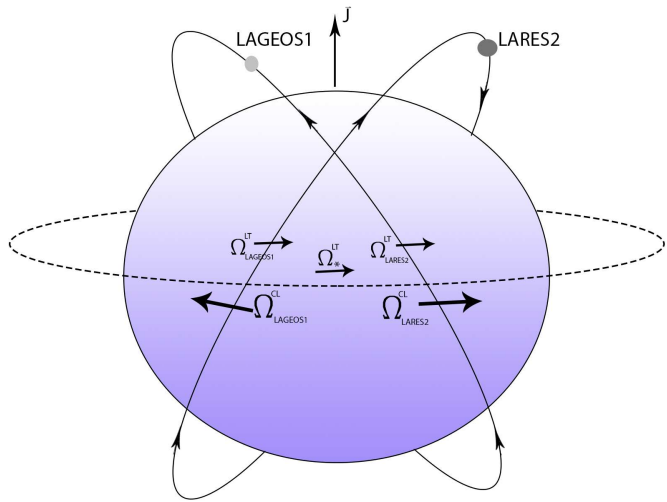}
\caption{The idea of the LARES 2 experiment (originally called LAGEOS 3) \cite{bib28,bib29} In the figure we show frame-dragging and the Newtonian precession of the nodes of two satellites with supplementary inclinations such as LARES 2 and LAGEOS (Supplementary inclinations: $i_{1}+i_{2}=180^\circ$).}
\label{fig:1}       
\end{figure}
Unfortunately, if the mass of the central body is not spherically symmetric, the nodal line of a satellite is also affected by a shift induced by Newtonian effects. When the Earth's gravitational potential is expanded in spherical harmonics, the {\itshape even zonal harmonics} are those of even degree and zero order \cite{bib27}. They represent axially symmetric deviations from spherical symmetry of the gravitational potential which are also symmetric with respect to the equatorial plane of the body. The main Newtonian secular drifts of the nodal longitude of a satellite arise from the Earth's even zonal harmonics. In particular, the largest node shift is by far due to the even zonal of degree two, J$_{2}$ (the Earth's quadrupole moment). The even zonal harmonics are extremely well measured by a number of techniques. Nevertheless even their tiny uncertainty produces a systematic bias in the measurement of frame-dragging. Even a tiny relative uncertainty of 10$^{-7}$ or less in the quadrupole moment corresponds to an uncertainty in the nodal rate comparable in order of magnitude to the frame-dragging effect. To eliminate all the errors due to the even zonal harmonics, in 1984-1989 we proposed the LAGEOS 3 satellite ([24,25,26,27] see also: [29,30] and [31, 32]). LAGEOS 3 is now called LARES 2. The idea of the LARES 2 space experiment is shown in Fig. 1. Whereas the Earth even zonal harmonics produce a shift of the nodal line of a satellite that is equal and opposite for two satellites with supplementary inclinations and equal semimajor axis, frame-dragging produces a shift of the nodal line that is in the same sense of rotation of the Earth for both satellites, independently of their inclination. (Supplementary inclinations: $i_{1}+i_{2}=180^\circ$.)
 Thus, by adding the measured residual nodal shifts of two such satellites, we can completely eliminate the components due to the errors in the Earth's even zonal harmonics while doubling the frame-dragging effect.

\section{A brief history of the tests of frame-dragging }

In this section we report a brief summary of the state-of-the-art tests of frame-dragging and in particular of the LAGEOS, LARES and LARES 2 space missions. In 1984-1989 a new laser-ranged satellite called ``LAGEOS 3'', identical to the LAGEOS satellite (launched in 1976 by NASA) was proposed with orbital parameters identical to those of LAGEOS but a supplementary inclination, that is with inclination i = 70.16$^{o}$ and semimajor axis = 12270 km. The orbital eccentricity was required to be nearly zero (for LAGEOS it is equal to 0.0048). At that time the state-of-the-art Earth gravity field determination was GEMT1. The main scientific objective of the LAGEOS 3 satellite was to be
a measurement of frame-dragging, discussed in the previous section.
Beginning in the 1960s, the Gravity Probe B space mission was developed in the USA with the goal of a 0.1\% test
of frame-dragging. Gravity Probe B was launched in 2004. But, due to unexpected systematic errors affecting the
gyroscopes, the final result of its test of frame-dragging achieved an accuracy of only approximately 20\%, much less stringent than expected \cite{bib37}. In 2004-2010, using the data of LAGEOS, LAGEOS 2 (a laser-ranged satellite almost identical to LAGEOS, launched in 1992 by the ASI and NASA) and the vastly improved new gravity field determinations from the Space Geodesy mission GRACE, measurement of frame-dragging achieved  an accuracy of approximately 10\% \cite{bib9,bib10,bib14,bib15,bib39,bib40}. GRACE is an outstanding space mission that has improved the knowledge of the Earth gravity field by several orders of magnitude \cite{bib41,bib42}.

In 2012 the new laser-ranged satellite LARES was successfully launched by ASI with the new ESA launcher VEGA, mainly built by AVIO and ELV. LARES’ main goal is a measurement of frame-dragging with a few percent accuracy. In 2016, using LARES, LAGEOS and LAGEOS 2 and the Earth gravity field determinations from GRACE, frame-dragging was measured with an accuracy of about 5\% (\cite{bib8} see also \cite{bib16}). GRACE, while still providing data, it is operating now almost a decade beyond its planned lifetime. A new GRACE Follow-On space mission, scheduled for launch in 2017, will continue to improve the accuracy of measurement of the Earth gravity field and its variations.

\section{Error Sources in The LARES 2 Experiment}

In this section, we present an error analysis of the gravitational and non-gravitational perturbations that will affect the LARES 2 space mission and we conclude that LARES 2 will allow a measurement of frame-dragging with an accuracy of a few parts in a thousand.

We consider the four main sources of systematic errors in the LARES 2 experiment: orbital injection errors and the effect of the even zonal harmonics, the effect of non-zonal Earth harmonics and tides, solar and albedo radiation pressure, and thermal drag \cite{bib28,bib29,bib30,bib31,bib32,bib33,bib34,bib35,bib36}. Each of these is estimated to produce an error of approximately 0.1\%, which when root-sum-squared (RSS) added results in an overall 0.2\% systematic error estimate. The discussion below also describes some smaller perturbing effects that contribute insignificantly to the RSS error. Monte Carlo simulations and covariance analyses, which fully support the present error budget, are presented in the next paper [7].

\subsection{Orbital Injection Errors}

The main error source in the originally proposed LAGEOS 3 experiment was due to the satellite orbital injection errors and in particular to the injection error in the inclination. No launch vehicle can achieve an orbit for LAGEOS 3, or LARES 2, which has an inclination {\itshape exactly} supplementary to that of LAGEOS and has exactly the same semimajor axis. Thus the cancellation of even-zonal errors would not be perfect, and that would introduce a systematic error in the measurement of frame-dragging proportional to the uncertainty in the quadrupole moment times the deviation of the inclination from that exactly supplementary to the one of LAGEOS.

Today, thanks to the space missions GRACE and to the forthcoming GRACE Follow-On, the knowledge of the Earth gravity field has dramatically improved. The knowledge of the Earth spherical harmonics expansion has improved by about three orders of magnitude with respect to the older, 1987, GEMT1 Earth gravity field.  The two groups responsible for the GRACE Follow-On mission (CSR of the University of Texas at Austin and GFZ of Potsdam) have estimated that at the time of the observations of GRACE Follow-On (with launch scheduled in 2017), the relative mean uncertainty in the quadrupole moment will be approximately   10$^{-8}$ (that is, a mean  uncertainty of about 0.5 $\cdot$ 10$^{-11}$  in its mean value of $\cong$ 0.484 $\cdot$  10$^{-3}$ ); see the next paper on LARES 2 \cite{bib7}. A 3-sigma orbital
 inclination injection error (about 0.15 degrees) of the new launch vehicle VEGA C induces an error of only about 10$^{-3}$  in the frame dragging due to non-perfect supplementary inclination of LARES 2 with respect to LAGEOS. A
smaller uncertainty would be induced by the injection error in the semimajor axis of LARES. So the orbital injection error will be:
\begin{center}
\textbf{Orbital injection error $\cong$ 0.1\%}
\end{center}

\subsection{Non-zonal Earth harmonics and Tides}

The non-zonal Earth spherical harmonics do not produce secular effects on the node, however they produce periodic nodal shifts that may introduce a bias in the measurement of frame-dragging \cite{bib27}. Thanks to GRACE and GRACE Follow-On, the spherical harmonics relevant to the frame-dragging measurement (the ones with the lowest degree) have an accuracy for the LARES 2 experiment improved by two or three orders of magnitude compared to the 1989 analysis with GEMT1.

The Earth tidal models have also drastically improved compared to those available for the older 1989 analysis, and it is expected that they will further improve in the near future. Furthermore, in our recent paper \cite{bib8} we demonstrated
a new method that allows us to dramatically reduce the error due to periodic effects with known
periods on the orbits of LAGEOS and LAGEOS 2 (tides, non-even zonal
harmonics and some non-gravitational perturbations). Therefore the bias in the measurement of frame-dragging with LARES 2 and LAGEOS due to non-even zonals and tides will at a level of about 0.1\%:
\begin{center}
\textbf{Non-even zonals and tides error $\cong$ 0.1\%}
\end{center}

\subsection{Non gravitational perturbation}

In the following subsections we estimate the errors due to direct solar radiation pressure, satellite eclipses, albedo, thermal drag, thermal drag and satellite eclipses, and particle drag \cite{bib30,bib31,bib33,bib34,bib44,bib45,bib46,bib47,bib48,bib49}.

\subsection{Direct solar radiation pressure}

The thrust of a satellite due to solar radiation pressure is proportional to the solar constant (radiation energy from the Sun per unit of time per unit of area), to the Cr of the satellite, a parameter depending on the reflection properties of the satellite and on its geometry, and on the cross-sectional area-to-mass ratio, A/M. The shift of the node of a satellite with small orbital eccentricity (such as LAGEOS) is proportional to its eccentricity. The solar constant is extremely well measured, and so is the Cr  of LAGEOS, thanks to 40 years of observations of its orbit. Furthermore A/M for LAGEOS is approximately equal to 0.007 cm$^{2}$/g and is the smallest of any satellite except LARES, and the orbital eccentricity of LAGEOS is only 0.0048. Thus the systematic error in the measurement of frame-dragging on the node of LAGEOS due to the uncertainty in modeling solar radiation pressure estimated to be at the level of less than 0.1\%.

For LARES 2, in order to reduce the systematic error due to radiation pressure (both direct solar and albedo, see below), as well as the systematic errors due to the other non-gravitational perturbations, we will increase its area-to-mass ratio by a factor of at least 1.5 with respect to LAGEOS, and also to decrease its orbital eccentricity. The reduction of its A/M ratio can be achieved by keeping its mass as much as possible equal to that of LAGEOS and by reducing its size. According to some preliminary calculations, depending also on the VEGA C launch constraints, we propose the mass of LARES 2 to be 350 kg, or more, and its diameter to be about 40 cm. And we propose an eccentricity of LARES 2 of 0.0025 or less. This will gain a factor of about 3, with respect LAGEOS, to reduce the nodal shift of LARES 2 due to solar radiation pressure. This will lead to a systematic error in the measurement of frame-dragging with LAGEOS and LARES 2, due to solar radiation pressure, that will be less than 0.1\%.

\subsection{Satellite eclipses}

An error in the modeling of the solar radiation pressure is that due to the satellite eclipses by the Earth. However the boundary of the shadow region generated by the Earth is well measured and included in our orbital estimators: GEODYN (NASA), EPOSOC (GFZ) and UTOPIA (CSR-UT), thus we do not expect a significant contribution from this source, as long as the step-size of the numerical orbital integrator is kept small enough to capture all crossings accurately. In conclusion, based on past extensive analyses, and by also considering the improved cross-sectional area to mass ratio of LARES, we will have an error in our experiment of less than 0.1\% due to direct solar radiation pressure coupled with satellite eclipses.

\subsection{Albedo}

Albedo produces the radiation pressure due to the sunlight reflected by the Earth surface. By reducing the cross-sectional area-to-mass ratio of LARES 2, according to previous extensive calculations of the effect of the albedo on LAGEOS and LAGEOS 3, and with the past improvements in the albedo models, we would then have a systematic error in the measurement of frame-dragging with LARES 2 and LAGEOS due to albedo radiation pressure of about 0.1\%
\begin{center}
\textbf{Albedo error $\cong$ 0.1\%}
\end{center}

\subsection{Thermal drag}

The electromagnetic radiation from the Sun and the radiation from the Earth each instantaneously heat one hemisphere of LAGEOS. Because of the finite heat conductivity of the body, there is an anisotropic distribution of temperature on the satellite and thus there is an anisotropic flux of energy, $\sim$ T$^{4}$, and momentum from the surface of the satellite, giving rise to its acceleration. If the satellite is spinning fast enough, the anisotropy in the satellite temperature distribution is mainly latitudinal. This is called the Yarkovsky or Yarkovsky-Schach effect.

Another thermal thrust effect was discovered on LAGEOS by David Rubincam, the Earth-Yarkovsky or Yarkovsky-Rubincam effect \cite{bib44,bib45}. Infrared radiation from the Earth is absorbed by the LAGEOS retro-reflectors. Therefore, due to the retro-reflector's thermal inertia and due to the past rotation of the satellite (today LAGEOS is almost rotationally at rest), there was a LAGEOS latitudinal temperature gradient. The corresponding thermal radiation caused an acceleration with an along-track component opposite to the satellite motion. An extensive analysis of the thermal forces acting on LARES is now under scrutiny by Richard Matzner and our LARES-team colleagues of the University of Texas at Austin \cite{bib47,bib48}.

Although today the LAGEOS satellite is rotationally almost at rest with respect to inertial space (i.e. it is spinning extremely slowly), Rubincam has calculated that there will still be an along-track component of the Yarkovsky-Rubincam effect but no out-of-plane component. This out-of-plane component is the one potentially responsible for the nodal drag. Therefore that drag is today substantially null for LAGEOS. In regard to the solar Yarkovsky effect, if the orbit does not intersect the Earth's shadow, the force is directed away from the Sun, and will simply add to the effect of the direct solar radiation pressure and thus will be accounted for by our estimation of the C$_{r}$ of LAGEOS (see our forthcoming paper analyzing the thermal drag on LARES 2, paper IV \cite{bib49}).

\subsection{Thermal Drag and Satellite Eclipses}

When the orbit of LAGEOS intersects the Earth's shadow we still have the Yarkovsky effect. Due to the satellite's thermal inertia, the satellite takes time to cool down in the shadow and then takes time to warm back up after it exits the shadow. Thus, in general, we will get radial, out-of-plane, and along-track thermal drag forces, with the Yarkovsky effect depending on the Sun-satellite orbital geometry. The LAGEOS satellite is in the shadow of the Earth for less than 1/20 of its orbital time; the cross-sectional area-to-mass-ratio, A/M, for LARES 2 will be about 1.5 times smaller than that of LAGEOS, and LARES 2 should be injected into orbit with a small spinning rate. The use of an alloy with high thermal conductivity will further reduce the thrust due to thermal anisotropy. Given the satellite characteristics, it will be possible to use a copper alloy that has a thermal conductivity about 3 times higher than the one of the tungsten alloy used for LARES. Finally, LARES 2 should be made of a single sphere (as LARES) and not by assembling different parts (as LAGEOS). With these facts, based on previous extensive analyses of the Yarkovsky effect, we estimate \cite{bib49} the systematic error due to the coupling of the Yarkovsky effect with the eclipses by the Earth at the level of 0.1\%:
\begin{center}
\textbf{Thermal Drag Error $\cong$ 0.1\%}
\end{center}

\subsection{Particle drag}

Particle drag is responsible for an along-track acceleration of a satellite. In the case of LAGEOS this leads to a very small decrease of its semimajor axis of approximately 1 millimeter per day. However, according to well-known theorems of celestial mechanics, the node of a satellite does not change due to particle drag and this result applies also to the case of a rotating atmosphere. Therefore, even by considering a rotating atmosphere at the LAGEOS altitude and by considering variations in its density, we obtained a negligible nodal shift of LAGEOS and LARES 2 due to particle drag \cite{bib33}.

\subsection{Measurement Errors of the Orbital Parameters}

Satellite Laser Ranging (SLR) provides the position of the LAGEOS satellites with a precision of the normal points
of less than a millimeter. This precision is certainly enough to accurately measure the shift of the node of LAGEOS due to
frame-dragging (almost 2 meters per year). However, the other LAGEOS and LARES 2 orbital elements must also be measured with high accuracy. Whereas the semimajor axis and the eccentricity of the LAGEOS satellites are measured with enough accuracy, the measurement error in the inclination of both LAGEOS and LARES 2 may introduce an error in the measurement of frame-dragging (the corresponding uncertainty in the determination of the Lense-Thirring effect using LAGEOS and LARES 2 will be induced by the error in the measurement of the inclination and not by the uncertainty in the modeling of the inclination, i.e. the uncertainty in the prediction of its behavior). The measurement uncertainty in the inclination of LAGEOS is mainly due to the atmospheric refraction. However, the refraction errors in the measurement of the inclination drop significantly at elevations above 20 degrees and most of the laser-ranging observations are indeed at elevations above 20 degrees. Therefore, by keeping in the data analysis only those observations corresponding to elevations above 20 degrees, we can reduce the error in the measurement of the inclination to a small level that, when propagated into the nodes of LAGEOS and LARES 2, would correspond to an error of no more than 0.1\% of frame-dragging.

\section{Final error budget}

The RSS error budget of the LARES 2 experiment is then summarized in Table 1.
\begin{table}[h!]
  \centering
  \begin{tabular}{|l|l|}
     \hline
     Source of Error & Estimated error \\ \hline
     Injection Error and Even Zonal Harmonics & $\cong$ 0.1\% of frame-dragging \\\hline
     Non-zonal harmonics and tides & $\cong$0.1\% of frame-dragging \\\hline
     Albedo &$\cong$ 0.1\% of frame-dragging \\\hline
     Thermal Drag and Satellites Eclipses & $\cong$0.1\% of frame-dragging\\\hline
     Measurement Error of the LAGEOS and LARES 2 Orbital Parameters & $\cong$ 0.1\% of frame-dragging \\\hline
       &   \\
     \textbf{Total RSS Error} & \textbf{$\cong$0.2\% of frame-dragging} \\
     \hline
   \end{tabular}
  \caption{Relative errors in the LARES 2 measurement of frame dragging}\label{tab1}
\end{table}

We have not included here the errors, listed above, which are smaller than 0.1\% since they would only affect the next decimal digit of the $\cong$ 0.2\% RSS total error. This LARES2 error budget was  confirmed by the Monte Carlo simulations and covariance analyses shown in the next paper \cite{bib7}.

\section{Mission requirements, orbital parameters and satellite characteristics}

Here we finally report the required LARES 2 orbital parameters and satellite characteristics.

The LARES 2 satellite must have the same semimajor axis of LAGEOS but orbital inclination supplementary to that of LAGEOS. The eccentricity must be zero in order to minimize the non-gravitational perturbations. The maximum deviations from the optimal orbital parameters are dictated by the requirement of a 0.1\% error due to the non-perfect elimination of the uncertainties due to the even zonal harmonics.

Thus, the proposed LARES 2 orbital parameters must be:
\begin{itemize}
  \item Semimajor axis 12270  km $\pm$ 20  km

  \item Inclination of LARES 2 = 70.16$^{o}$ $\pm$ 0.15$^{o}$ (supplementary to that of LAGEOS)

  \item Orbital eccentricity: between 0 and 0.0025

\end{itemize}

These values are fully compatible with the VEGA C 3-$\sigma$ injection accuracies.

The mass and radius of the satellite must be chosen to minimize the cross-sectional-to-mass ratio, thus minimizing the non-gravitational perturbations. However, there is a limitation on the mass (between 350 kg and 400 kg) that can be injected to an altitude of 5900 km using the VEGA C launch vehicle. The higher value in the proposed range is the preferred one because it will further reduce the effect of the non-gravitational perturbations thus reducing the final error budget. The proposed radius is about 20 cm.

\section{LARES 2 and space geodesy}

Satellite Laser Ranging (SLR) now provides uniquely the most accurate definition of the origin of the International Terrestrial Reference Frame (ITRF) and has an equal share with VLBI in the definition of its scale. This is accomplished using the two LAGEOS SLR targets, designed for cm-level geodesy and not for today’s goals of 1 mm and 0.1 mm/y. The precise orbits of the GNSS satellites are referred to reference frames based on the ITRF frame.
The ITRF will realize the United Nations (UN), ``Global Geodetic Reference Frame
(GGRF)", the reference standard for all applications. The UN adopted this proposal only recently (UN resolution A/RES/69/266, \url{http://www.unggrf.org}). The repercussions, however, will last for decades to come. LARES, the third SLR target with its improved specifications, enhanced the status quo ante. But even with LARES we are far from having the ideal ``SLR Constellation'', compared to the GNSS ones. The ideal constellation would comprise a number of LARES-class spacecraft. The development of several LARES-like satellites in combination with the existing LAGEOS, LAGEOS 2 and LARES satellites will deliver a large number of tracking opportunities with multiple SLR targets above the horizon of each ground station a few times per day (Figure 2).

\begin{figure}
\centering
 \includegraphics[width=0.640\textwidth]{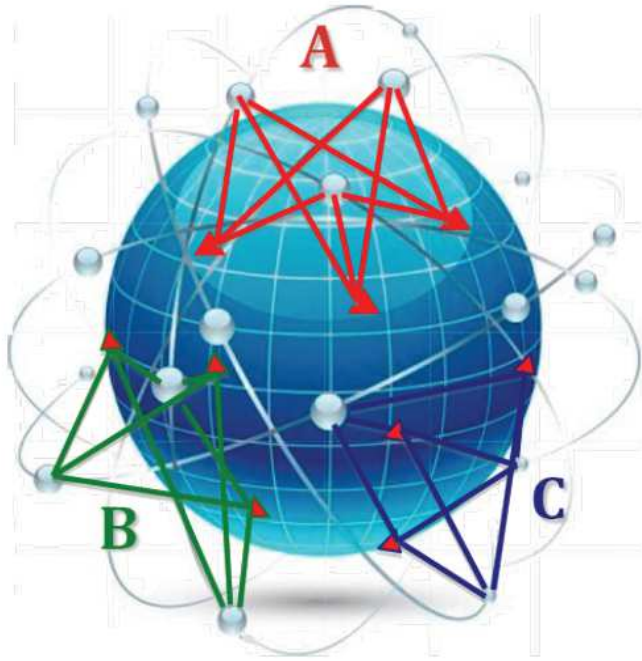}
\caption{Conceptual s/c distribution for a hypothetical future SLR Constellation with three regional sub-networks A, B, C.  }
\label{fig:2}       
\end{figure}

The construction
and launch of LARES 2 will provide the fourth satellite of the SLR Constellation. With LARES 2 added to the
constellation we can achieve much better results compared to today's situation. The analysis of almost three years of
data from LAGEOS 1 and LAGEOS 2, since the LARES launch (2012-2014), compared to those from a similar time
span before LARES was launched, indicates that the mean site position improved by 17\% while the Earth  Orientation Parameters (Polar motions and Length-of-Day) improved by 21\% \cite{bib51,bib52}. The addition of LARES 2 and more LARES-class targets will improve the current knowledge by more than the square root law, since it is mainly the geometry of the problem that improves, something that cannot be described by simple statistical arguments. In addition to the optimal determination of the ITRF and subsequently its optimal distribution to users via the GNSS orbits, the estimation of other terrestrial geophysical parameters, such as the secular evolution of low degree zonal harmonics, the terrestrial tides, elastic properties of Earth, etc., will also benefit from the enhancement of the SLR space segment.

\section{Summary and conclusions}

The Gravity Probe B experiment has published in
2011 a measurement of frame-dragging with accuracy of about 20\%. With the LARES space experiment we have so far achieved a 5\% measurement of frame-dragging [8]; this test could
eventually be improved to reach a final accuracy of about 2\%. With LARES 2 we could obtain a 0.2\% test of
frame-dragging. Thus the LARES 2 satellite would provide a factor 10 improvement over the best test of frame-dragging
that could be achieved with the satellites orbiting today. Monte Carlo simulations and covariance analyses presented
in the next paper have fully confirmed the present error analysis (Paper II [7]).

%
%
%
%
%
%
%

%
%

%
%

\end{document}